\documentclass[authoryear, 12p]{elsarticle}
\usepackage{amstext}
\usepackage{amssymb}
\usepackage{graphicx}
\usepackage{lineno}

\makeatletter

\usepackage{hyperref}

\journal{Icarus}

\begin{document}

\begin{frontmatter}

\title{Minerals detection for hyperspectral images using adapted linear
unmixing: LinMin}

\author{Schmidt Fr{\'e}d{\'e}ric$^{1,2}$, Maxime Legendre$^{3}$, St{\'e}phane Le Mou{\"e}lic$^{4}$}

\address{$^{1}$ Univ Paris-Sud, Laboratoire IDES, UMR8148, Orsay, F-91405;
(frederic.schmidt@u-psud.fr) $^{2}$ CNRS, Orsay, F-91405.Universit{\'e}
Paris Sud $^{3}$ IRCCyN (CNRS UMR 6597), Ecole Centrale Nantes, France,
$^{4}$ LPGN, CNRS UMR 6112, Universit{\'e} de Nantes, France}

\begin{abstract}
Minerals detection over large volume of spectra is the challenge addressed
by current hyperspectral imaging spectrometer in Planetary Science.
Instruments such OMEGA (Mars Express), CRISM (Mars Reconnaissance
Orbiter), M$^{3}$ (Chandrayaan-1), VIRTIS (Rosetta) and many more,
have been producing very large datasets since one decade. We propose
here a fast supervised detection algorithm called LinMin, in the framework
of linear unmixing, with innovative arrangement in order to treat
non-linear cases due to radiative transfer in both atmosphere and
surface. We use reference laboratory and synthetic spectral library.
Additional spectra are used in order to mimic the effect of Martian
aerosols, grain size, and observation geometry discrepancies between
reference and observed spectra. The proposed algorithm estimates the
uncertainty on ``mixing coefficient'' from the uncertainty of observed
spectra. Both numerical and observational tests validate the approach.
Fast parallel implementation of the best algorithm (IPLS) on Graphics
Processing Units (GPU) allows to significantly reduce the computation
cost by a factor of \textasciitilde{}40.
\end{abstract}


\end{frontmatter}


\section{Introduction}

Various methods have been proposed to detect surface chemical species
(minerals, ice) on large dataset of hyperspectral images. Supervised
methods (knowing the spectra of the chemical species you want to detect)
are widely used, for instance : band ratio techniques (\citep{Poulet_OMEGA-MnxMap_JGR2007,Ehlmann_PhylosolicateMars_Nature2011,Carter_Automatedprocessingplanetary_PaSS2013}),
linear unmixing (\citep{Combe_MELSUM_PSS2008,Themelis_OMEGAhyperspectral_PSS2012}),
wavelet based detection (\citep{Schmidt_Wavanglet_IEEETGRS2007,Gendrin_ClassifHyperspectral_JGR2006}),
correlation based detection - Spectral Angle Mapper (\citep{Kruse1993}).
Other linear non-supervised techniques (estimating the spectra directly
from the scene) have been proposed, using Independent Component Analysis
ICA (\citep{2005LPI....36.1623F,Erard_ICAvirtis_JGR2009}) or
Bayesian methods under constrained (\citep{Moussaoui_JADE-BPSS_Neurocomp2008,Schmidt_ImplementationBPSS_TGRS2010}).
Also some algorithms only extract the endmember spectra such: Pixel
Purity Index (\citep{1995Boardman}), N-FINDR (\citep{Winter_NFINDR:algorithm_1999}),
or graph-based segmentation (\citep{Gilmore_SuperpixelHyperspectral_JGR2011})
.

The linear mixture is still valid in non-linear intimate mixture,
but with significant difference between retrieved ``mixing coefficient''
and actual abundance (\citep{Mustard_Photometry_JGR1994,Mustard_nonlinear_JGR1998}).
Nevertheless, linear unmixing is satisfactory for mineral detection
(\citep{Combe_MELSUM_PSS2008,Themelis_OMEGAhyperspectral_PSS2012}).
It has the significant advantage to deal with complex mixture of a
large variety of candidate minerals (see fig. \ref{fig:32ReferenceSpectra}),
whereas band ratio methods fails, due to the lack of defined reference
``continuum'' wavelength channel. In such situation each wavelength
may sample an absorption band of a particular mineral and no reference
wavelength can be found. If band ratio's are adapted to detect minerals
over flat spectra, methods using the entire wavelength channels (such
linear unmixing) are adapted to detect minerals over a complex variety
of background. Specifically, we draw attention that the band ratio's
of the CRISM summary products are only relevant in case of a pure
mineral detection but may be irrelevant in case of a mixture. We propose
here to address the challenge to detect one mineral type alone or
a mineral type in an assemblage, from a very diverse potential endmember
spectra dataset.

Linear unmixing is relatively simple enough to provide a fast implementation.
Furthermore supervised linear unmixing has the advantage that its
interpretation is directly and automatically provided in terms of
mineralogical class on the bulk hyperspectral image. Dimensionality
reduction methods, such unsupervised algorithms or endmember extractions
are not required in linear unmixing. Also linear unmixing does not
require manual interpretation and identification of spectra.

Previous linear unmixing algorithms have proposed to optimize the
reference spectral library in order to solve positivity (\citep{Combe_MELSUM_PSS2008}),
forcing for sum-to-one constraints on the mixing coefficients (\citep{Roberts_MappingChaparralin_RSoE1998}).
\citep{Roberts_MappingChaparralin_RSoE1998} discusses the benefits
of constraining the number of endmembers from the spectral library.
They have shown that the solutions with a mixture of two endmembers
have less overlap that three endmembers mixtures, probably due to
the dimension space. In our article, we focus on the interest of both
positivity and sum-to-one constraints, that allow to estimate the
optimum of all possibility of mixtures. Without positivity and sum-to-one
constraints, singularities may arise due ``linear dependance'' of
endmember spectra (\citep{Roberts_MappingChaparralin_RSoE1998,Dalton_Exogeniccontrolssulfuric_PaSS2013}).
We show that most recent algorithms, including both positivity and
sum-to-one constraints, handle this difficulty, thus simplifying tremendously
the treatment.

Our aim is to show that additional spectra (flat, slope) minimize
the differences between reference laboratory spectra and actually
observed spectra, by reducing the effect of grain size and continuum
level (photometry, surface roughness, compaction). Some differences
between observations and reference endmembers may still be present
due to variations in composition, texture,... The selection of the
reference spectra database requires always special care and depends
on the scientific questions addressed, but at least this database
is explicitly known.

Once detection has been performed, time consuming inversions must
be applied in order to retrieve quantitative estimate of surface properties
(exact mineralogy, abundances, grain size, porosity, ...) as proposed
by different techniques such MGM (\citep{Sunshine_MGM_JGR1993,Kanner_LimitMGM_Icarus2007})
or radiative transfer inversion (\citep{Doute_PSPC_PSS2007,Poulet_MaficOMEGA1theorie_Icarus2009}).
Also the particular mineral type, for instance Al/Fe or Mg/Fe phyllosilicate,
must be addressed by additional work using specific spectral library.

We propose here to extend the linear supervised detection technique
using a new set of constraint (positivity and sum to unity) and a
new set of additional spectra in order to improve the ``mixing coefficient''
estimation. We also introduce a new renormalization with the goal
to fit the signal, even in the case of very low signal to noise ratio. 

We propose to call LinMin our strategy of using: 
\begin{enumerate}
\item linear unmixing technique under constraints (positivity, sum to unity/sum
lower than unity)
\item additional spectra made of flat and slope (also possibly cosine functions)
\item renormalization by the variance/covariance noise matrix
\item estimation of the error on the ``mixing coefficient
\end{enumerate}
In order to validate the LinMin approach, we test the detection limits
on synthetic data simulating a regolith mixture, the grain size effect
and the Martian aerosols effect. We also validate the method on actual
hyperspectral dataset of OMEGA, CRISM and M$^{3}$. Quantitative estimate
of the detection limits are computed.

\section{Method}

\subsection{Linear unmixing}

The linear mixing model of hyperspectral reflectance is usually written
as:

\begin{equation}
X=A.S+E\label{eq:LinearMixing}
\end{equation}

with the collection of observed spectra $X$ ($M\times N_{\lambda}$
matrix), the reference spectra $S$ ($N\times N_{\lambda}$ matrix),
the unknown mixing coefficients $A$ ($M\times N$ matrix) and the
additive noise error $E$ ($M\times N_{\lambda}$ matrix), assumed
to be gaussian with zero mean. $N_{\lambda}$ is the number of wavelength,
$N$ the number of reference spectra, $M$ the number of observed
spectra.

The unmixing problem then consists in the estimation of the mixing
coefficients $A$ that minimize the error $E$. Considering the least
squares error minimization, it is written:
\begin{equation}
\min_{A\in\Re^{M\times N}}F(A)\quad\textrm{with}\quad F(A)=\sum_{m=1}^{M}\sum_{\lambda=1}^{N_{\lambda}}(X_{m\lambda}-(A.S)_{m\lambda})^{2}\label{eq:UnconstrainedProblem}
\end{equation}

The unconstrained solution $A=X.S^{T}.\left(S.S^{T}\right)^{-1}$
has been used the in previous detection methods (\citep{1995Boardman,Combe_MELSUM_PSS2008})
but the retrieved mixing coefficient $A$ may be negative. One solution
is to test all combination of reference spectra $S$ in order to keep
only positive case (\citep{Roberts_MappingChaparralin_RSoE1998,Combe_MELSUM_PSS2008}),
but the computation cost is very high. In addition, such case is more
sensitive to degeneracies in comparison to the constrained problem
(see paragraph \ref{sub:Reference-spectra-database}).

\subsection{Linear unmixing under constraint}

The previous formulation is not sufficient to describe physical constraints
on the mixing coefficients. Indeed, the mixing coefficients must satisfy
non-negativity : 
\begin{equation}
\forall m\in\{1,\ldots,M\}\;\forall n\in\{1,\dots,N\}\; A_{mn}\geq0\label{eq:NonNegativityConstraint}
\end{equation}
and sum-to-one constraints:
\begin{equation}
\forall m\in\{1,\dots,M\}\;\sum_{n=1}^{N}A_{mn}=1\label{eq:SumToOneConstraint}
\end{equation}

\emph{Non-Negative Least Squares} (NNLS) algorithms(\citep{Lawson_1995,bro1997fast})
aim at solving the problem \ref{eq:UnconstrainedProblem} subject
to constraint \ref{eq:NonNegativityConstraint}. \emph{Sum-to-one
Constrained Least Squares} (SCLS) methods \citep{settle1993linear}
solve the same problem with constraint \ref{eq:SumToOneConstraint}.
Eventually, \emph{Fully Constrained Least Squares} (FCLS) algorithms
(\citep{Heinz_FullyConstrainedLinear_TGRS2001,dobigeon2008semi})
solve the problem \ref{eq:UnconstrainedProblem} subject to both constraints
\ref{eq:NonNegativityConstraint} and \ref{eq:SumToOneConstraint}.
In this article, two FCLS methods are considered: (i) IPLS based on
primal dual interior point optimization (\citep{Chouzenoux_Algorithmeprimal-dualde_GRETSI2011})
that benefits from GPU implementation (\citep{Chouzenoux_FastConstrainedLeast_JSTARS2013}),
(ii) BI-ICE in the bayesian framework (\citep{Themelis_NovelHierarchicalBayesian_IToSP2012}). 

As stated in \citep{Chouzenoux_FastConstrainedLeast_JSTARS2013},
the convexity of the criterion F is sufficient to establish the convergence
of IPLS when constraints \ref{eq:NonNegativityConstraint} and \ref{eq:SumToOneConstraint}
are considered. Hence, the unmixing estimation can be performed even
with correlated reference spectra, such in fig. \ref{fig:4-Additional-spectra}.
Next sections will describe the results on actual data.

\subsection{Measurement uncertainty consideration}

In some cases, the level of uncertainty is known and can be modeled
by a gaussian probability density function with zero mean and the
covariance $C$ ($N_{\lambda}\times N_{\lambda}$ matrix). $C^{-1}$
being a symmetric positive definite matrix, it can be factorized using
the Cholesky decomposition: 
\begin{equation}
C^{-1}=L.L^{T}
\end{equation}
 with $L$ lower triangular. Thus, noting $X=[X_{1},\dots,X_{M}]^{T}$
and $A=[A_{1},\dots,A_{M}]^{T}$, the least squares criterion becomes:

\begin{equation}
\min_{A\in\Re^{M\times N}}F(A)\quad\textrm{with}\quad F(A)=\sum_{m=1}^{M}(X'_{m}-A_{m}.S')(X'_{m}-A_{m}.S')^{T}\label{eq:RenormalizedProblem}
\end{equation}
with $X'=X.L$ and $S'=S.L$. Thereby, the measurement uncertainty
is handled by modifying the observation matrix $X$ and the reference
spectra $S$ before the unmixing process.

An estimation of the error on the mixing coefficient A is directly
provided by the Hessian (\citep{Chouzenoux_FastConstrainedLeast_JSTARS2013}).

\subsection{Reference spectra database\label{sub:Reference-spectra-database}}

As stated in introduction, the reference spectra database depends
on the scientific goal of the detection. The total number of spectra
is limited to the number of wavelength. In the following, we used
32 spectra of the main minerals type proposed to be present at the
surface of Mars and the Moon (see fig. \ref{fig:32ReferenceSpectra}).
More endmembers can be used but with increasing computation time (\citep{Chouzenoux_FastConstrainedLeast_JSTARS2013}).

The linear unmixing problem (eq. \ref{eq:UnconstrainedProblem}) is
degenerated if the spectra in the database are linearly dependent.
Degeneracies can create false solution with low RMS error (\citep{Roberts_MappingChaparralin_RSoE1998,Combe_MELSUM_PSS2008}).
We estimate here the linear dependence of the database in order to
show that positivity constraint significantly decreases the appearance
of degeneracies. This test is an ``auto-fit'' of the reference spectra
database. For each spectra of the database $S_{n}$, we estimate the
abundance matrix $A'$, for a new set of reference spectra made of
all endmembers except $S_{n}$ ($S'=S\smallsetminus S_{n}$). We use
the unconstrained problem (eq. \ref{eq:UnconstrainedProblem}):

\begin{equation}
S_{n}=A'.S'+E
\end{equation}
and also the non-negative constrained problem (eq. \ref{eq:UnconstrainedProblem}
subject to constraint \ref{eq:NonNegativityConstraint}) to estimate
$A'$. 

\begin{figure}
\includegraphics[width=1\textwidth]{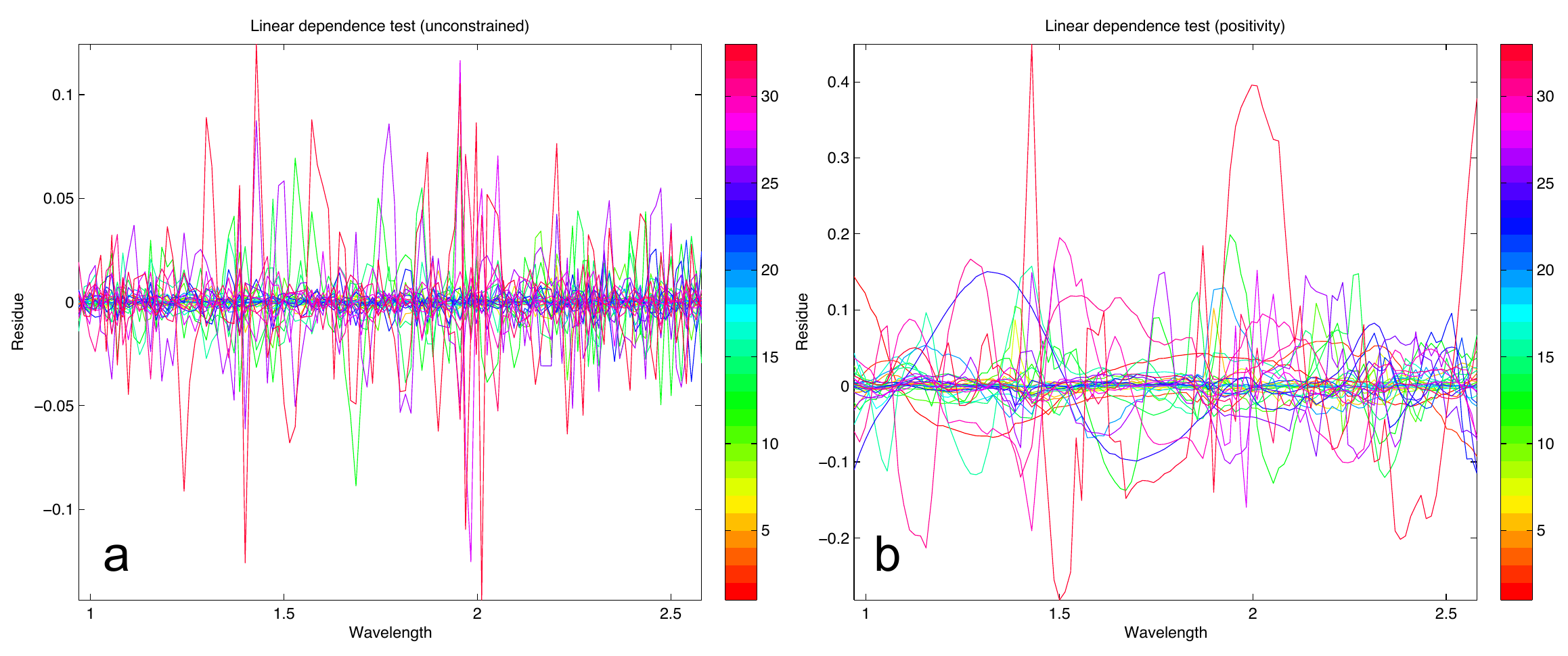}\centering\caption{Estimation of degeneracies of the reference database: (a) on left
without positivity constraints, (b) on right with positivity constraints.\label{fig:Estimation-of-degeneracies}}
\end{figure}

Figure \ref{fig:Estimation-of-degeneracies}, shows the residue $E$
between the real endmember $S_{n}$ and the best linear mixture of
all other endmember $A'.S'$, for all endmembers. The unconstrained
problem shows smaller residues than the positivity constrained problem.
In addition, the residues are unstructured (noisy) for the unconstrained
problem, in contradiction to the very structured residues (inverted
spectra) for the positivity constrained problem. This fact indicates
that the spectra $S_{n}-E$ is better fitted with a linear mixture
of false endmember $S'$instead of the real case. In conclusion, an
observed spectra with noise can likely create false detection in the
unconstrained problem but much less likely in the positivity constrained
problem. 

\subsection{Benefits of additional spectra}

In order to model the effect of shading, grain size and aerosols,
\citep{Combe_MELSUM_PSS2008} propose to add a positive and a
negative linear slope, and also a flat spectrum in order to correct
for the Mie scattering of the aerosols particles (see fig. 2 in \citep{Combe_MELSUM_PSS2008}).

\begin{figure}
\includegraphics[bb=0bp 0bp 500bp 420bp,clip,width=0.7\columnwidth]{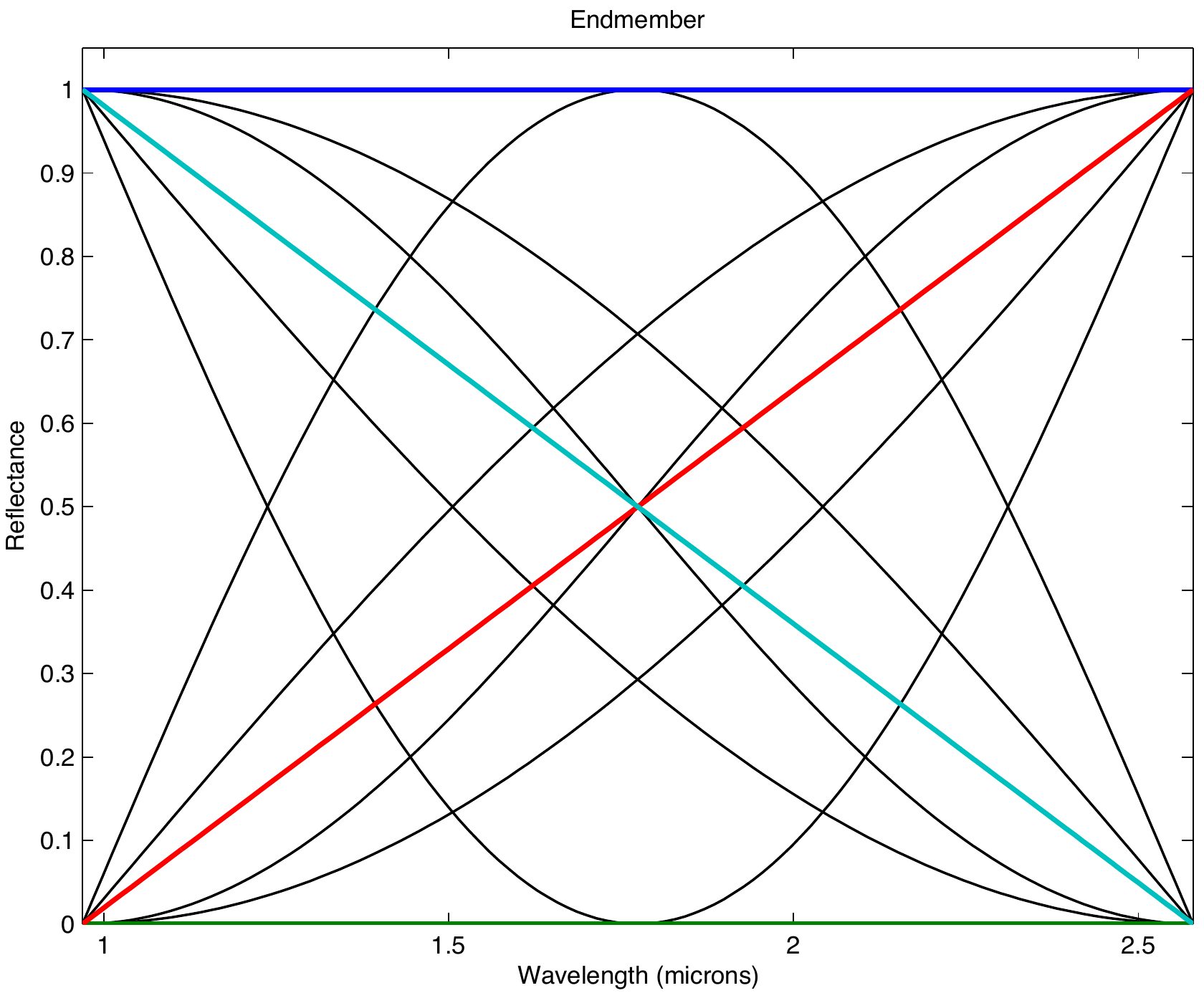}\centering\caption{thick color lines : 4 additional spectra to fit difference of level
and slope between $X$ and $S$. thin black lines : 8 cosine/sine
function to model the large scale difference between $X$ and $S$.
\label{fig:4-Additional-spectra}}
\end{figure}

This strategy can be extended in the case of linear unmixing under
constraints. Since mixing coefficients are positive and constrained
to one, the flat spectra split in two : flat near zero level and also
near 1 level (see fig. \ref{fig:4-Additional-spectra}). Those 4 additional
spectra allows to model the difference of level and slope between
$X$ and $S$.

In order to model more complex continuum shape such aerosols, we could
also add cosine function with periods x2 and x4 of the spectral domain
(\citep{Schmidt_LinearUnmixing_whispers2011_submission110,Schmidt_AccurateDataBaseMineral_EPSC2012}).
Similarly to Fourier transform, those cosine functions are orthogonal
and linearly independent, allowing unique mixing coefficient. In the
framework of linear unmixing under positivity and sum to one constraint,
12 additional spectra are required. Those 12 additional spectra allow
to model the large scale difference between $X$ and $S$, similar
to Fourier filtering. In particular the aerosols contribution that
peak at the wavelength similar to the grain size may be fitted by
large scale cosine. Figure \ref{fig:SyntheticFit} (b) show a pure
aerosol spectra (aot=100), that could be modeled by level/slope/sine/cosine
functions. Since the aerosols grain size may change, leading to an
unknown peak over the wavelength, our approach will be able to fit
it. 

Since those 4 (or 12) additional spectra are highly correlated, from
algorithmic point of view it may hard to estimate the mixing coefficient.
To our knowledge, the only algorithms that are able to estimate the
unmixing under constraints and high correlated reference spectra are
IPLS (\citep{Chouzenoux_Algorithmeprimal-dualde_GRETSI2011})
and BI-ICE (\citep{Themelis_NovelHierarchicalBayesian_IToSP2012}).
Next section will describe the results on actual data.

\section{Synthetic tests}

In order to test the effect of various discrepancies between observation
and reference spectra on detection accuracy, we propose to simulate
3 cases: linear random mixture at the surface without atmosphere,
linear random mixture at the surface observed trough a diffusive atmosphere,
grain size change.

\subsection{Surface mixture\label{sub:Surface-mixture}}

\begin{figure}
\includegraphics[bb=0bp 0bp 500bp 420bp,clip,width=0.7\columnwidth]{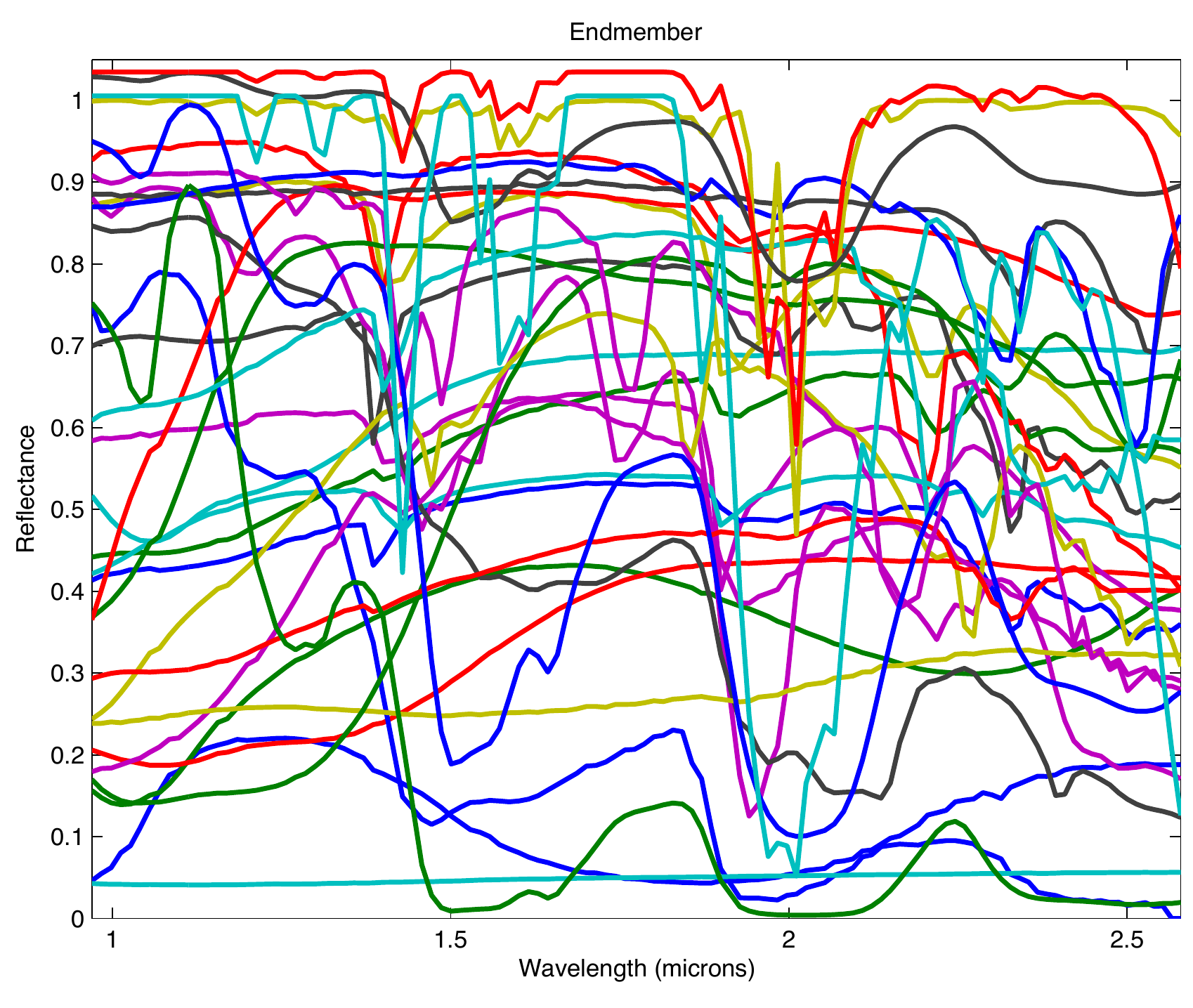}\centering\caption{32 Reference spectra of minerals, ice and atmospheric gas representing
major classes of contributions of surface spectra. The difficulty
to define continuum wavelength is due to the overlapping signatures
of all species. See Appendix for the names. }
\label{fig:32ReferenceSpectra}
\end{figure}

In order to simulate a realistic set of spectra, we create a random
set of 1000 binary mixture spectra from our 32 reference spectra database
(see fig. \ref{fig:32ReferenceSpectra}) from USGS catalog (\citep{Clark2003}),
the CRISM Analysis Tool (CAT), synthetic spectra from radiative transfer
model (\citep{Doute_reflectancemodel_JGR1998}). Each simulated
spectra are composed of:
\begin{itemize}
\item 90\% is a flat component at reflectance 0.35 in agreement with OMEGA
studies \citep{Vincendon_Marssurfacephase_PaSS2013} in order
to reproduce the low level and flatness of actual Martian spectra.
The flatness is also representative of the Moon or other planetary
surface.
\item 10\% of a random mixture of two over 32 reference spectra with random
uniform mixing coefficients, noted as $A_{0}$. For each endmember
i, there is \textasciitilde{}30 spectral mixture with non-null mixing
coefficient (noted $A^{positive}$) and \textasciitilde{}970 with
null mixing coefficient ($A^{false}$).
\end{itemize}

We add synthetic noise, simulating the noise level of a typical OMEGA
observation after gas correction. We estimate the noise covariance
matrix from dark current noise of ORB41\_1, transferred into the gas
corrected calibrated observation space. The noise has a wavelength-average
standard deviation of $1.3\times10^{-3}$. Other noise statistics
can be used, such MNF shift difference from ENVI software, but we
point the fact that OMEGA dark current is archived, estimating the
minimum noise statistics (excluding spike and other non-linear effects).

The estimated mixing coefficients are noted $A_{IPLS}$, $A_{BI-ICE}$
and $A_{IPLS,cov}$ for the renormalized problem. 

Examples of fits are shown in fig. \ref{fig:SyntheticFit} on left
with AOT=0.

\begin{figure}
\includegraphics[bb=0bp 0bp 500bp 420bp,clip,width=0.7\columnwidth]{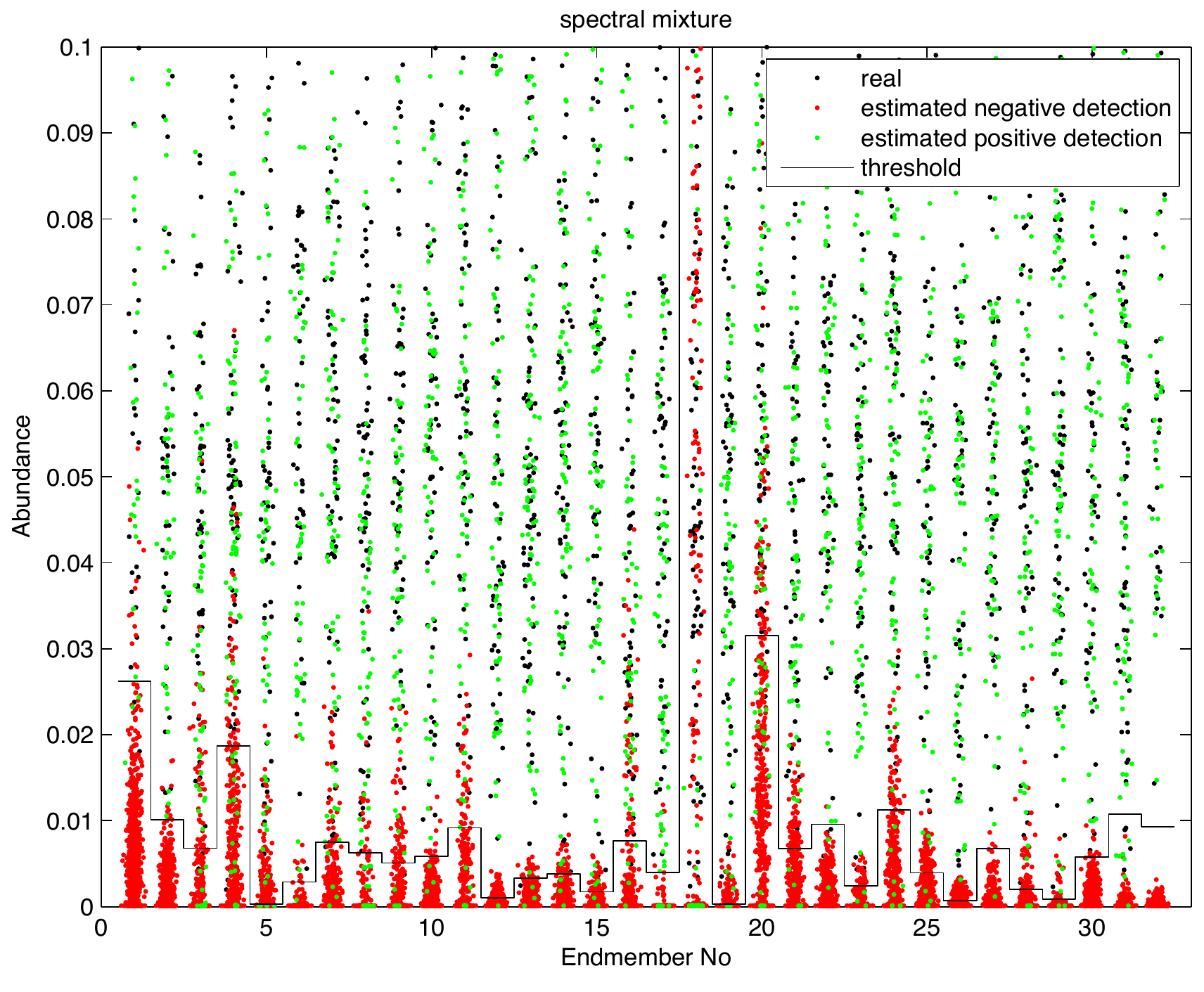}\centering\caption{Mixing coefficient for 32 spectra (see fig. \ref{fig:32ReferenceSpectra})
for the IPLS algorithm on 1000 binary mixture, solving the renormalized
unmixing problem (eq. \ref{eq:RenormalizedProblem}), using 36 endmembers
(32 spectra and the 4 additional spectra). black points: actual mixing
coefficient $A_{0}$, generated randomly. green points: estimated
mixing coefficient $A_{IPLS,cov}^{positive}$ in the case of positive
detection. red points: estimated mixing coefficient $A_{IPLS,cov}^{false}$
in the case of false detection. black line: best threshold $C_{IPLS,cov}^{i}$
to detect the endmember. Please note that artificial dispersion around
the endmember No has been added in order to increase the clarity of
the figure.}
\label{fig:ResultsAbundanceIPLS} 
\end{figure}

Figure \ref{fig:ResultsAbundanceIPLS} compares estimated $A_{IPLS,cov}$
and actual $A_{0}$ mixing coefficients, for the IPLS algorithm using
32 and the 4 additional spectra, on the 1000 binary mixture, with
the renormalized unmixing problem. Usually, $A_{IPLS,cov}$ for actually
present endmember ($A_{IPLS,cov}^{positive}$ in green), are well
separated from the absent endmember ($A_{IPLS,cov}^{false}$ in red),
showing that a simple threshold can be used to detect the endmember.
Endmember No 18 (Magnetite) is not detectable because there is no
difference between $A_{IPLS,cov}^{positive}$ and $A_{IPLS,cov}^{false}$
. Please note that $A_{IPLS,cov}^{positive}$ and $A_{0}^{positive}$
are very close, the mean absolute difference is 0.0142, one order
of magnitude lower than the mixing coefficients (up to 10\%), showing
that when linear mixing is true, the estimated mixing coefficients
are validated as real abundances. Also, the root-mean-square (RMS)
error is compatible with the level of noise ($1.2\times10^{-3}$).

In order to estimate the best threshold $C_{IPLS,cov}^{i}$, avoiding
the maximum number of false detection but keeping the maximum positive
detection, we propose the following formulation, for each endmember
$i$:

\begin{equation}
C_{IPLS,cov}^{i}=\frac{1}{2}\left(mean(A_{IPLS,cov}^{positive})-2.std(A_{IPLS,cov}^{positive})+mean(A_{IPLS,cov}^{false})+6.std(A_{IPLS,cov}^{false})\right)\label{eq:Threshold}
\end{equation}
Figure \ref{fig:ResultsAbundanceIPLS} shows these threshold for all
endmembers. It confirms that some endmembers can be detected with
a larger detection limits than other, depending on their relative
similarities. Endmember No 18 (Magnetite) is not detectable because
its spectral shape is too close to a combination of the 4 artificial
spectra, because its spectra is featureless.

Actual positive detection and false detection rates without atmosphere
are shown in figure \ref{fig:ResultsAbundanceDetectionRates} at the
AOT=$10^{-3}$ for IPLS with/without renormalization, for IPLS and
BI-ICE, for 32 or 36 or 44 endmembers. For all algorithms the 4 additional
spectra (called slope), clearly improve the classification over the
32 endmembers case (called no slope). Without renormalization, IPLS
shows the best results ($>$ 70\% of positive detection, $<$ 20\% of false
detection) over BI-ICE. With renormalization, the detection rates
is improved ($>$ 85\% of positive detection, $<$ 5\% of false detection)
for IPLS. For non-renormalized IPLS, the sum-to-one (sto) seems slightly
better than sum-lower-than-one (slo). 

The use of 12 additional spectra (including the sine/cosine), seems
to be less efficient that 4 additional spectra (83 \% of positive
detection, 5\% of false detection). It appears that the accuracy of
detection slightly decreases for some minerals with large scale feature
such olivines or pyroxenes. Since, those phases are major ones, we
decide to focus on the 4 additional spectra.

As a summary, the best algorithm to detect linear mixture is clearly
IPLS with renormalization and 4 additional spectra. Please note that
a different threshold may change the positive/false detection rates
but not change the relative accuracy of the classification. Estimating
the threshold for real case images is more difficult due to the possible
differences between observation and reference spectra. Please note
that our test uses the same spectra in the spectral library and in
the mixture.

\subsection{Aerosols}

\begin{figure}
\includegraphics[clip,width=1\textwidth]{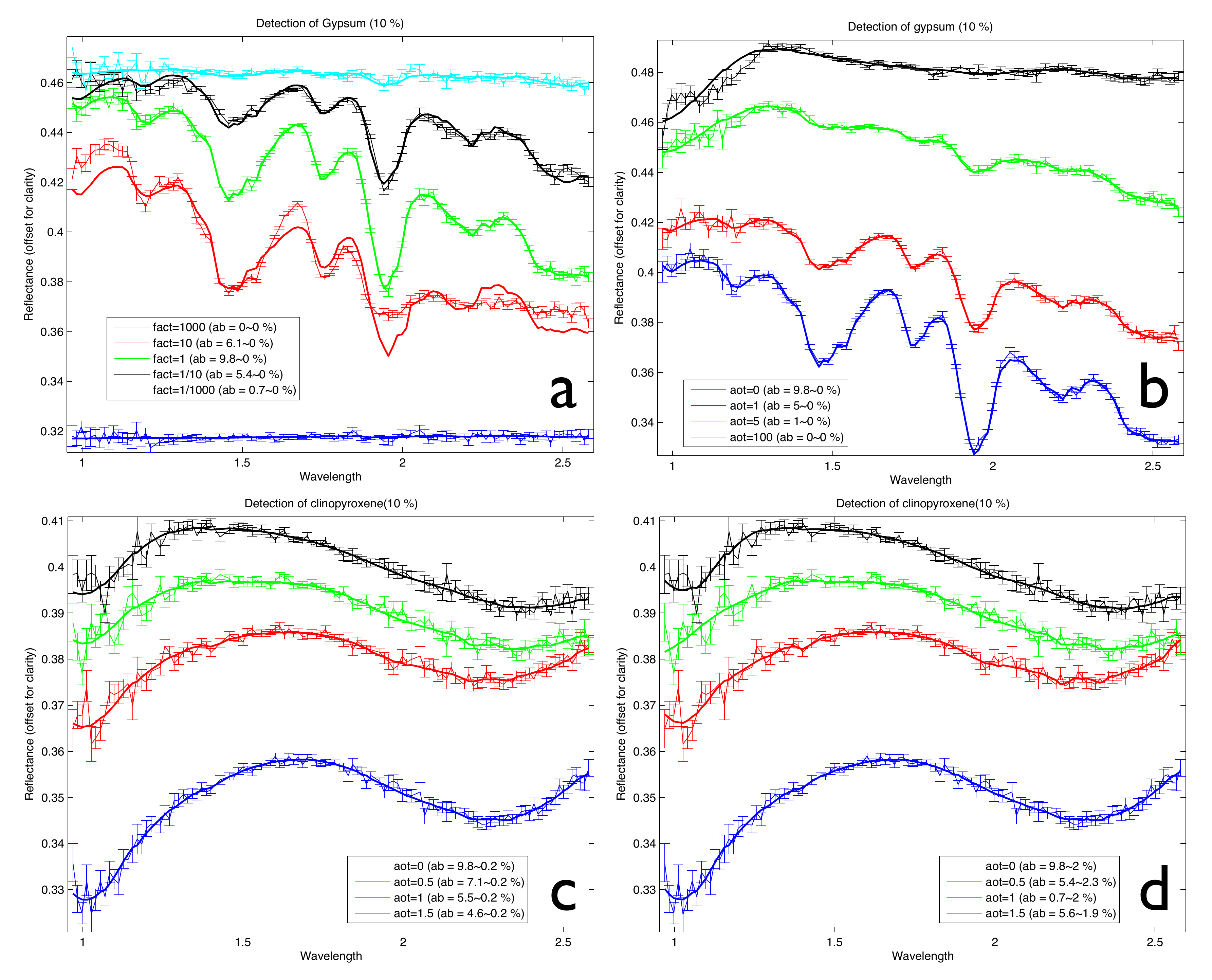}\centering\caption{Examples of results using the IPLS algorithm: a) and b) with renormalization
and 12 additional spectra (flat and slope) in the case of a pure gypsum
spectra (10\% in abundance). a) with different grain size factor,
from x1000 to x1/1000, b) with different atmospheric load from AOT=0
to AOT=100. ; c) and d) with renormalization in the case of diopside
(clinopyroxene). c) using 4 additional spectra (flat and slope), d)
using 12 additional spectra (flat, slope and cosine).}
\label{fig:SyntheticFit} 
\end{figure}

In order to test if our methodology is able to reduce the effect of
the non linear surface-atmosphere coupling due to aerosols, we modified
the previous surface spectra by non linear radiative transfer using
aerosol properties from \citep{Vincendon_DustAerosols_JGR2007}.
These authors propose a parameterization of the dust aerosol properties
in form of a single scattering albedo and a shape of optical thickness
as a function of wavelength, rescaled to a Aerosols Optical Thickness
at 1 micron (AOT). The reflectance of a semi-infinite aerosol media
is plotted in figure \ref{fig:SyntheticFit} b, for AOT=100. The simulation
uses the DISORT algorithm (\citep{Stamnes_DISORT_AO1988}) with
the bottom condition as lambertian surface with the previous surface
mixture spectra. We took into account the reflection and absorption
of the aerosols but also the multiple reflection between surface and
atmosphere. We used the following geometry, typical of hyperspectral
observation: emergence angle 0$^\circ$, incidence angle 76$^\circ$,
azimuth angle 60$^\circ$. Other optical properties and/or geometries
may be used but we assume that the chosen properties are reasonable
to simulate aerosols effects. We sample 104 values of AOT (Aerosols
Optical Thickness):
\begin{itemize}
\item The value 0 for no aerosols, identical to surface only (noted $10^{-3}$
in the log scale graphs)
\item 99 value in a logarithm from 0.01 to 4.6 using $-\log(\frac{100-[1:99]}{100})$. 
\item Four values : 5, 10, 20 and 100 to sample the optically thick cases.
\end{itemize}
We obtain a set of 104 000 spectra simulating the space-borne Martian
observation from the top of the atmosphere. We also add the synthetic
OMEGA noise in order to simulate a realistic observation using the
same approach that previously described.

On a desktop with Dual Core at 2.53 Ghz with 4Go RAM memory, the typical
computation time to solve the non-normalized problem of eq. \ref{eq:UnconstrainedProblem}
on $M=104000$ spectra with $N_{\lambda}=110$, $N=36$ are : 3.3
h for BI-ICE, 4.5 min for IPLS. The estimated mixing coefficients
are noted $A_{IPLS}$, $A_{BI-ICE}$. Since IPLS is the fastest algorithm,
we also tried the solve the harder renormalized problem of eq. \ref{eq:RenormalizedProblem}
and found a computation time of 1.2h. We also use IPLS to solve the
renormalized problem using 12 additional spectra. The estimated mixing
coefficients are noted $A_{IPLS,cov}$. Also GPU implementation on
a TESLA C 2050 (448 core at 1.15 GHz) may solve the problem with a
computation time of\textbf{ }several 10s. The GPU implementation efficiency
dependance on spectral/spatial size and number of endmember is discussed
in \citep{Chouzenoux_FastConstrainedLeast_JSTARS2013}.

\begin{figure}
\includegraphics[bb=0bp 0bp 500bp 500bp,clip,width=0.7\columnwidth]{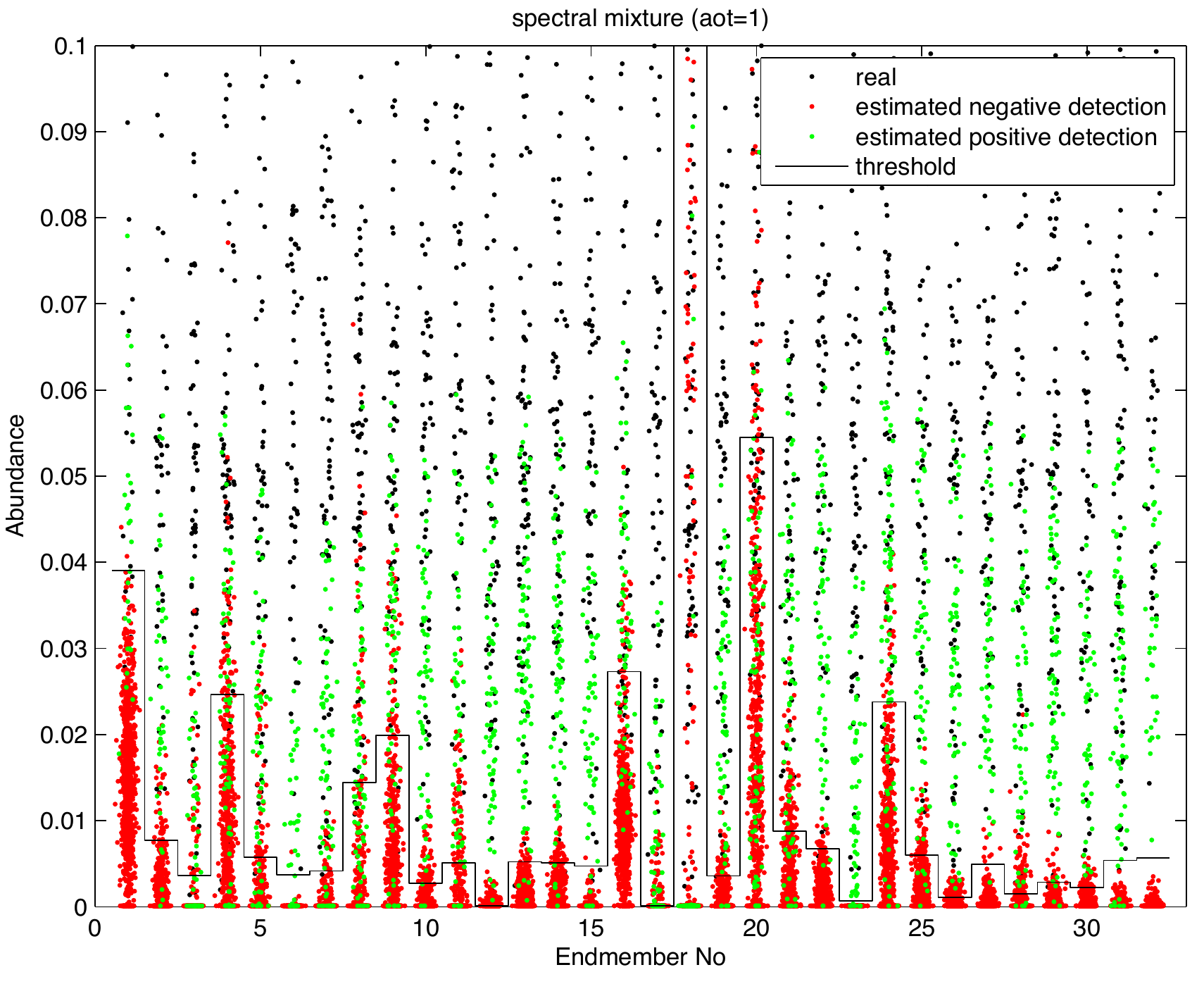}\centering\caption{Same fig. as \ref{fig:ResultsAbundanceIPLS} but with AOT=1 instead
of AOT=0.}
\label{fig:ResultsAbundanceIPLSaot1} 
\end{figure}

In order to illustrate the effect of AOT on the detection limits,
figure \ref{fig:ResultsAbundanceIPLSaot1} shows the same than figure
\ref{fig:ResultsAbundanceIPLS} but for AOT=1. The positive detection
mixing coefficients endmembers ($A_{IPLS,cov}^{positive}$ in green)
are well separated from the absent endmember ($A_{IPLS,cov}^{false}$
in red), showing again that a simple threshold can be used to detect
the endmember. Endmember No 18 (Magnetite) is still not detectable.
Please note that $A_{IPLS,cov}^{positive}$ and $A_{0}^{positive}$
are getting far, the mean absolute difference is 0.0310, showing that
when linear mixing is not true, the estimated mixing coefficient may
differ from the true abundances. The estimated mixing coefficients
$A_{IPLS,cov}^{positive}$ are systematically lower than the ``true''
$A_{0}^{positive}$ because of the effect of aerosols, removing the
minerals signature ( see fig. \ref{fig:SyntheticFit}).

\begin{figure}
\includegraphics[clip,width=1\columnwidth]{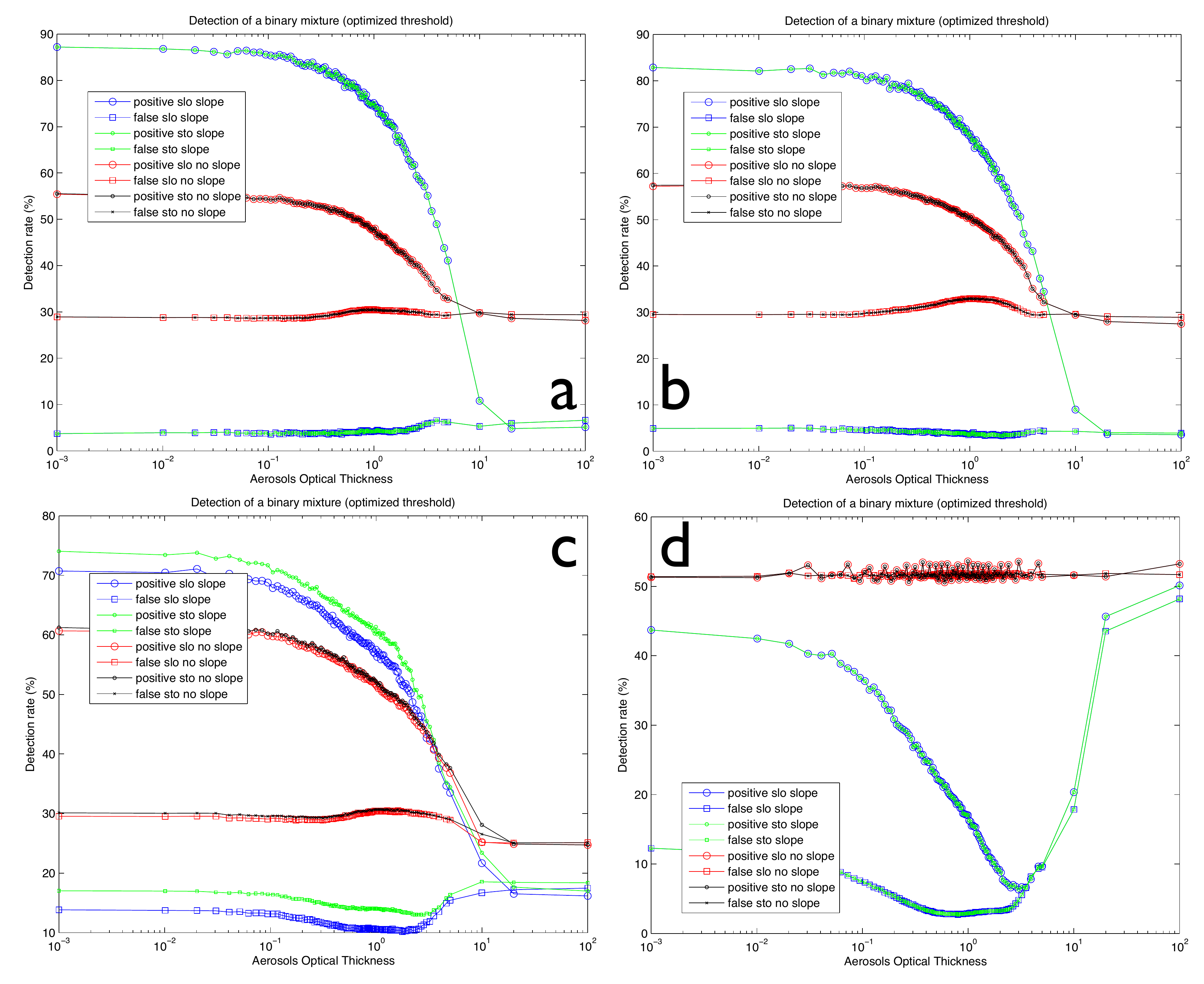}\centering\caption{Positive detection rate (circle) and false detection rates (square)
with optimized threshold as a function of AOT for (a) IPLS with renormalization
and 36 endmembers (b) IPLS with renormalization and 44 endmembers
(c), IPLS without renormalization and 36 endmembers (d) BI-ICE without
renormalization and 36 endmembers. For each algorithm, the positive/false
detection rates are computed using 32 endmember spectra only (no slope),
or 36/44 endmember spectra (slope). IPLS has an option of sum-lower-than-one
(slo) or sum-to-one (sto) that is not relevant for other algorithms.
The case without aerosols is plotted at AOT=$10^{-3}$.}
\label{fig:ResultsAbundanceDetectionRates} 
\end{figure}

Results on actual positive detection and false detection rates are
shown in figure \ref{fig:ResultsAbundanceDetectionRates} as a function
of AOT. The positive detection rates of all algorithms drop for AOT$>10^{-1}$
. For IPLS with renormalization, the positive detection rates is 70
\% at AOT=$10^{0}$ and 10 \% at AOT=$10^{1}$, showing again that
it is the best algorithm. IPLS with renormalization is also the only
providing a constant low false detection rate ($<$5 \%), independent
of the AOT. The RMS error for IPLS with renormalization and additional
spectra is still very low and compatible with the actual level of
noise (at maximum $1.7\times10^{-3}$ for AOT=$10^{2}$), showing
that the non-linearity of the radiative transfer due aerosols can
be fitted with our model. Examples of fits are shown in fig. \ref{fig:SyntheticFit}
b) demonstrating the effect of decreasing mixing coefficient as a
function of increasing AOT.

Figures \ref{fig:SyntheticFit} c) and d) show the effect of 4 additional
spectra (level and slope) or 12 additional spectra (level, slope,
cosine) on the detection of pure diopside. In the case of AOT=1 using
12 additional spectra, the mixing coefficient is estimated to be $0.7\pm2\%$,
showing a non-probable detection in spite of the actual 10\% abundance.
In the same case using 4 additional spectra, the mixing coefficient
is estimated to $5.5\pm0.2\%$, showing a clear positive good detection.
We argue that all endmember with large scale absorption bands could
have the same behavior and thus, we prefer to use 4 additional spectra
to detect them. Users focused on spectra endmember with small scale
features only, such phyllosilicate, could use the 12 additional spectra.

As a summary, the best algorithm to detect non-linear coupling of
a linear mixture under increasing aerosols content is IPLS with renormalization
and additional spectra. Please note that a different threshold may
change the positive/false detection rates but not change the relative
accuracy of the classification. Estimating the threshold for real
case images is more difficult due to the possible differences between
observation and reference spectra. Please note that our test uses
the same spectra in the spectral library and in the mixture. 

\paragraph*{Comparison with band ratio}

\begin{figure}
\includegraphics[width=1\textwidth]{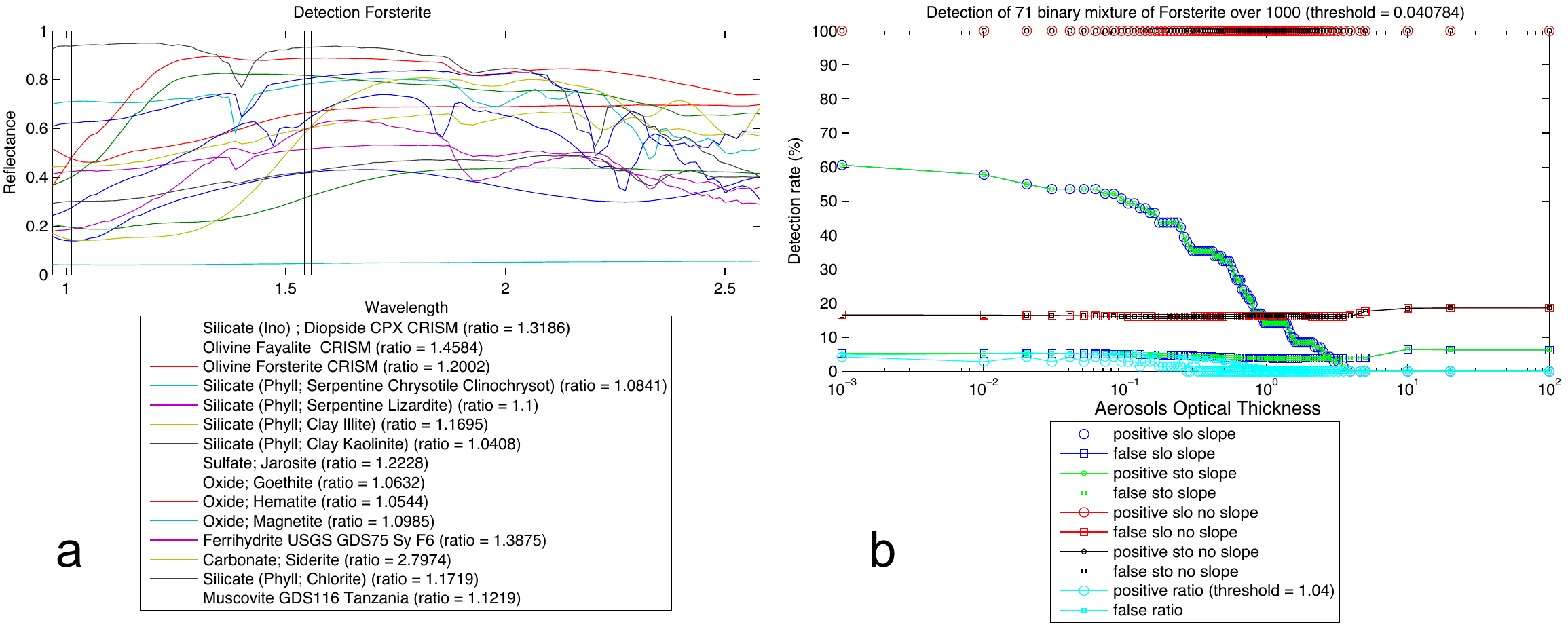}\centering\caption{Detection of forsterite using band ratio and LinMin strategy using
IPLS with renormalization and 44 endmembers. Only 71 mixtures over
1000 contains forsterite. (a) Detection rates and false detection
rates as a function of AOT (b) random examples of false detection
using band ratio ($>$1.04), but absence of detection with LinMin (mixing
coefficient $\sim10^{-5}$) \label{fig:Detection-of-forsterite}}
\end{figure}

In order to estimate the accuracy of those results with the mostly
used technique: band ratio, we compute the 1 microns band spectral
parameter to detect forsterite (\citep{Poulet_OMEGA-MnxMap_JGR2007,Ody_Globalinvestigationolivine_JGRP2013}),
called OSP1 (\citep{Ody_Globalinvestigationolivine_JGRP2013}),
in our dataset. Figure \ref{fig:Detection-of-forsterite} a) presents
the pure endmember with a ratio greater than the defined threshold
(1.04) (\citep{Poulet_OMEGA-MnxMap_JGR2007,Ody_Globalinvestigationolivine_JGRP2013}).
This result shows the phenomena of false detection that could occur
in the presence of other endmembers, indicating that band ratio is
relevant to detect the presence of one single mineral feature vs no
feature but it is not able to handle detect mineral against a large
dataset of mineral. This is due to the difficulty to define a ``continuum''
wavelength, as shown in figure \ref{fig:32ReferenceSpectra}.

Figure \ref{fig:Detection-of-forsterite} b) presents the detection
of forsterite of the 1000 synthetic examples as a function of AOT,
for the OSP1 band ratio and IPLS with different constraints. The band
ratio seems to be have difficulties to detect the presence of forsterite
in the 71 cases of mixture, because only 5\% are detected. IPLS is
able to detect up to 60\% of the mixture with forsterite with the
use of additional spectra, with a small range of false detection ($<$5\%),
similar to OSP1. In the case of IPLS without additional spectra, the
false detection rate is very large (\textasciitilde{}20\%) and independent
of the AOT, indicating wrong results. In this case, IPLS performs
a better detection with additional spectra.

As a summary, in case of diverse spectra and mixture, one single band
ratio is not able to detect the current spectra. This difficulty may
be tackled by defining several band ratios but it is more difficult
with an increasing number of candidate spectra. The LinMin strategy
performs significantly better with additional spectra.

\subsection{Grain size}

We choose the first 26 minerals spectra (see Appendix) with following
parameter to retrieve the optical index of the grain material using
the inverse Shkuratov theory (\citep{Shkuratov_ModelReflectance_Icarus1999}):
real optical constant=1.7, volume fraction filled by particles=0.8,
grain size=100 microns. This approach has been widely used in hyperspectral
data analysis (\citep{Poulet_ComparativeRadiativeTransfert_Icarus2002,Doute_PSPC_PSS2007,Poulet_MarsPetrologyMafic_Icarus2009}).

We then regenerate the reflectance spectra using the direct Shkuratov
theory but with 9 different grain size factor x1000, x100, x10, x5,
x1, x$\frac{1}{5}$, x$\frac{1}{10}$, x$\frac{1}{100}$,
x$\frac{1}{1000}$ of the original 100 microns. We obtain a set
of 234 synthetic spectra.

In order to simulate a realistic observation we use the same approach
that previously described for the surface mixture using flattening
of the spectra and OMEGA noise addition (see section \ref{sub:Surface-mixture}).

We do not claim that all reference spectra have been recorded at a
grain size of 100 microns so that the generated dataset has a determined
accurate grain size. Nevertheless, this manner allows us to generate
synthetic spectra simulating a grain size factor change. 

Since IPLS with renormalization is obviously the best algorithm, we
only tested this approach for the grain size. We used again the same
threshold strategy explained earlier in eq. \ref{eq:Threshold}. Here
positivity constraint only may be relevant in the case of higher grain
size than the endmember, since the signature are stronger and may
imply mixing coefficient larger than 1.

\begin{figure}
\includegraphics[clip,width=1\columnwidth]{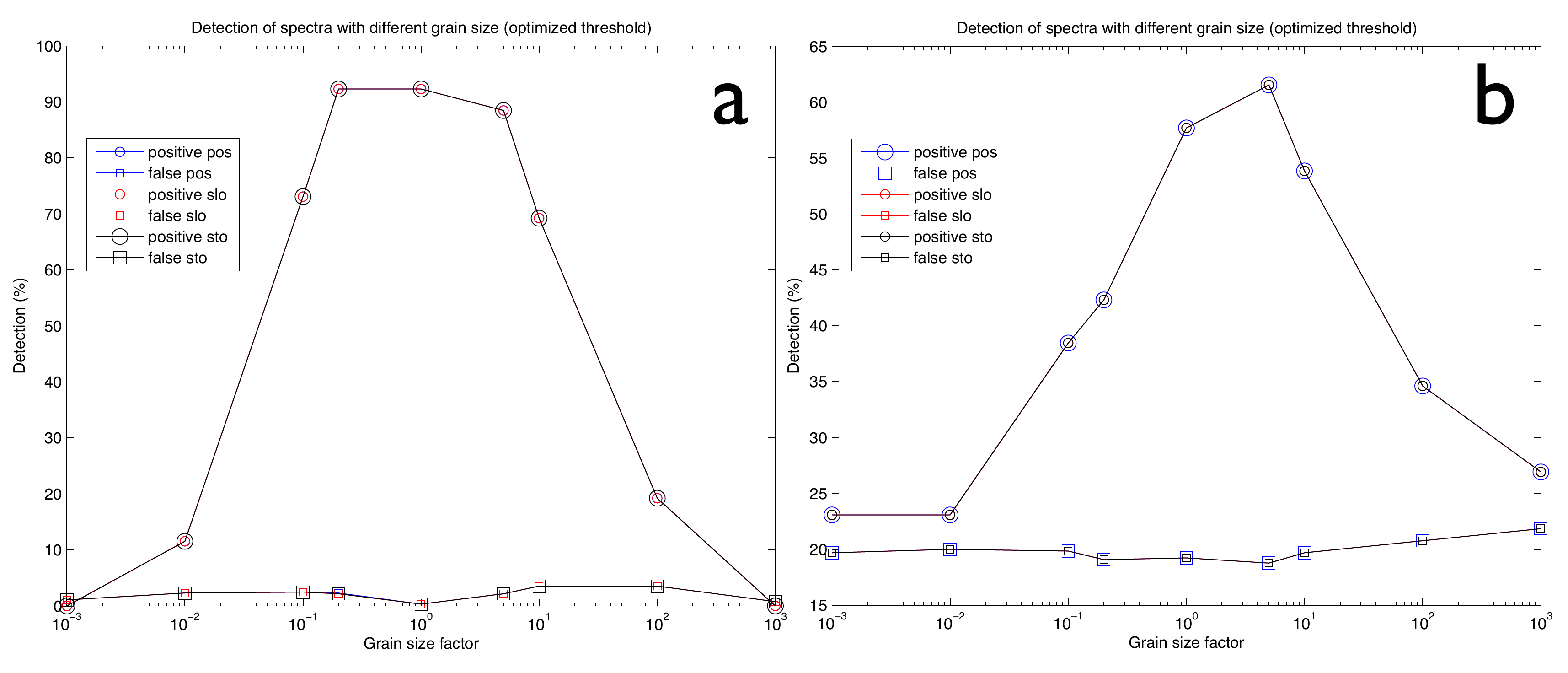}\centering\caption{Positive detection rate (circle) and false detection rates (square)
with optimized threshold as a function of grain size factor for IPLS
with renormalization: (a) using 36 endmember spectra (using additional
flat and slope endmember), (b) using 32 endmember spectra only (no
slope). IPLS has options of constraints: sum-lower-than-one (slo),
sum-to-one (sto) and positivity only (pos).}
\label{fig:ResultsAbundanceDetectionRatesGrainSize} 
\end{figure}

Positive and negative detection rates are plotted in fig. \ref{fig:ResultsAbundanceDetectionRatesGrainSize}
for 36 or 32 spectra, for three constraints of IPLS : sum-lower-than-one
(slo), sum-to-one (sto) and positivity only (pos). First of all, the
detection is nearly perfect ($>$90\%) for the grain size factor of x1,
x5 and x1/5 using additional 4 spectra. The false detection rate is
very low ($<$2\%). For 32 spectra, the positive detection is around
60\% for factor x1 to x10, and false detection is around 20\%. Using
the 4 additional spectra clearly improves the detection rates but
slo, sto and pos are equivalent in this case. For a factor of x$\frac{1}{10}$
to x10, the detection rate is higher than 65\%. It then decreases,
moving away from factor x1. The detection rates are quite symmetrical
in respect to increasing and decreasing grain size, with a slightly
lower detection rates for the increasing grain size. Examples of fits
are shown in fig. \ref{fig:SyntheticFit} demonstrating the effect
of decreasing mixing coefficient as a function of increasing/decreasing
grain size factor. This plot clearly shows that increasing grain sizes
are more difficult to handle with our strategy because the error between
estimated spectra and true one is higher. The main solution is to
provide the endmembers with grain size higher than expected.

The RMS error is still very low and compatible with the actual level
of noise (at maximum $2.1\times10^{-3}$), showing that the main non-linearity
effects due grain size factor can be fitted with our model.

As a summary, the best endmembers to detect mineral species in case
of a non-linear grain size effect again with the 4 additional spectra.
Please note that a different threshold may change the positive/false
detection rates but not change the relative accuracy of the classification.
Estimating the threshold for real case images is more difficult due
to the possible differences between observation and reference spectra.
Please note that our test uses the same spectra in the spectral library
and in the mixture. Intimate mixture may be generated from the Shkuratov
theory but the detection limits may be forecast from this discussion,
knowing that the largest grain size species is spectrally dominant
(as computed for ices in \citep{Schmidt_Wavanglet_IEEETGRS2007}
for instance). 

\section{Real case}

We propose to test our automatic algorithm to three different real
case from OMEGA, CRISM and M$^{3}$ instrument. For all cases, we
will compare our results with other detection from the literature,
mainly based on band ratio followed by ratio of the region of interest
spectra over a manually selected ``flat'' spectra. For all cases,
we do not test that the exact minerals types could be detected since
many equivalent spectra can be identified (clays, olivine, pyroxene,
...). The names are the names of the material from laboratory spectra.
For instance ``goethite'' is an example of iron oxide. Since all
mineral inside a class are spectrally close (like for iron-oxides),
it may be not possible to distinguish the precise mineral. We decided
to keep magnetite since it has been detected on Mars (\citep{Chevrier_ReviewMineralogyMars_PSS2007})
but estimated mixing coefficients are not relevant due to the featureless
spectra. More accurate detections are possible by adapting and improving
the endmember spectra database. All maps, spectral fits of the maximum
mixing coefficient and average of each compounds are available in
supplementary material.

From previous section, IPLS seems the best algorithm to solve the
LinMin problem so that we will exclusively use it on the real case
hyperspectral images.

The detection of minerals requires different conditions of good behavior
of the solution:
\begin{itemize}
\item The maximum mixing coefficient of the mineral should be higher than
a certain threshold. This condition permits to focus on the main spectral
component but component with smaller contribution can be present.
We define the best compromise threshold for each observation due to
noise level uncertainties. 
\item The error on mixing coefficient should be lower than the mixing coefficient,
in order to have a significant detection.
\item The RMS should be lower than the noise, basically estimated at 10
times the dark current noise.
\end{itemize}

\subsection{OMEGA}

We select the observation ORB422\_4 of Observatoire pour la Min{\' e}ralogie,
l'Eau, les Glaces et l'Activit{\' e} (OMEGA) onboard Mars Express
(MEx) (\citep{Bibring_OMEGA_ESA_SP-v1_2004}) of Syrtis Major
as a test case because this single cube contains well identified areas
with very strong signatures of mafic minerals (pyroxenes, olivines)
and phyllosilicates (\citep{Mustard_OlivinePyroxen_science_2005,Combe_MELSUM_PSS2008,Ehlmann_in-siturecordof_GRL2012}).
The data cube has been radiometrically calibrated and the atmospheric
gas transmission has been empirically corrected using the volcano
scan method (\citep{Erard_MarsSpectra_Icarus1997,Langevin_SulfatesPolar_science_2005}).
We estimated the noise covariance matrix on the calibrated and atmosphere
corrected cube from the dark current. We then applied the renormalized
IPLS algorithm using 36 spectra and sum-lower-than-one constraint. 

\begin{figure}
\includegraphics[clip,scale=0.4]{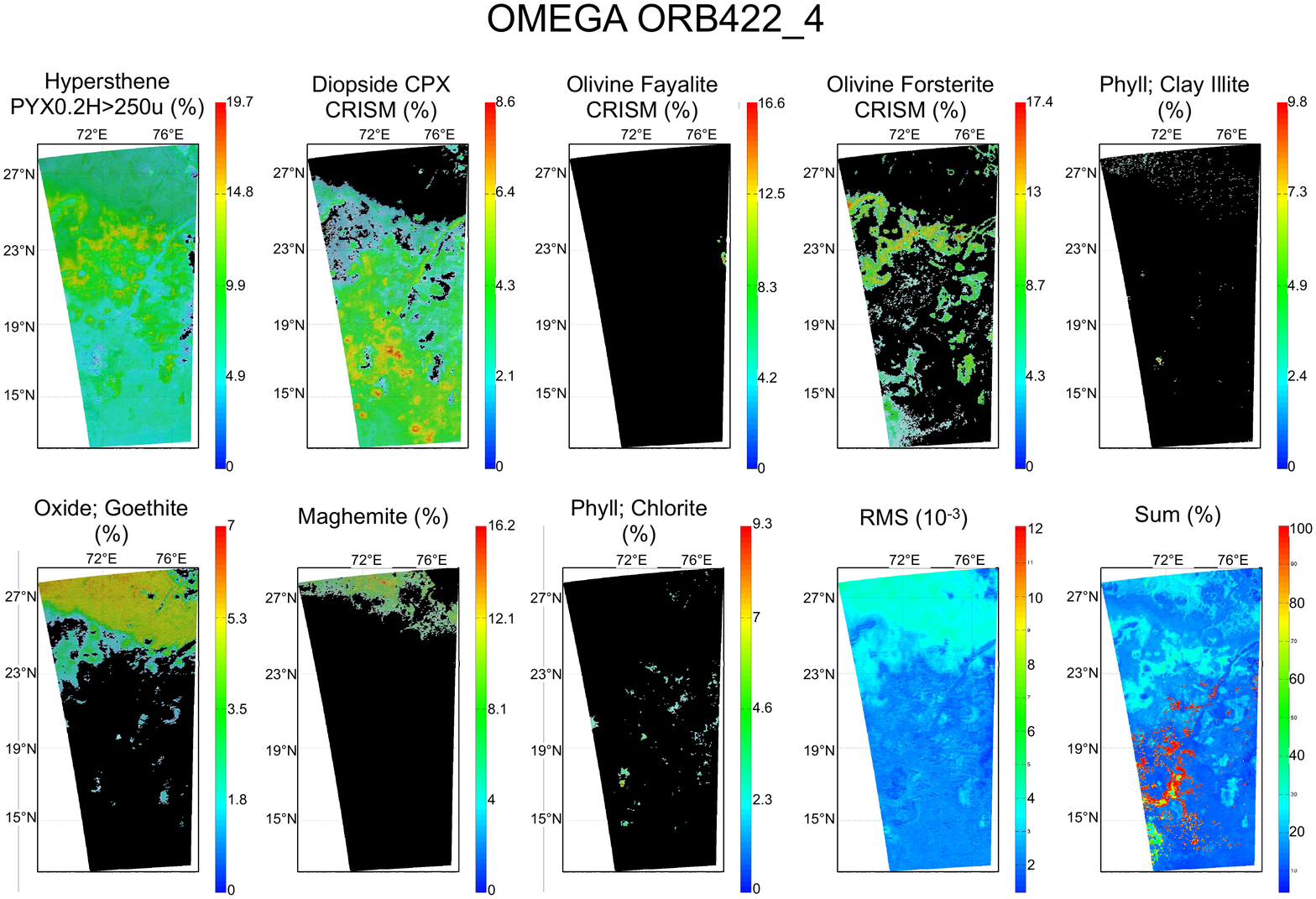}\centering\caption{Detection of 8 minerals over 36 spectra on OMEGA image ORB422\_4 of
Syrtis Major using IPLS in the hue-saturation-value color system.
The hue (color) represents the mixing coefficient. The saturation
(color or b/w) represents the error. The value (intensity of color
or b/w) represents the rms. Spectral mixing coefficient map are shown
with following conditions : (i) maximum mixing coefficient $>$ 5\% ,
(ii) error on mixing coefficient $<$ mixing coefficient, and (iii) RMS
$<$ 10x the dark current noise (see text). Pyroxenes, olivines, phyllosilicates
and oxides are detected and the corresponding ``mixing coefficient``
are mapped (color refer to the online version of the article).}
\label{fig:OMEGAdetectionSyrtisMajor} 
\end{figure}

The mixing coefficient results, plotted in figure \ref{fig:OMEGAdetectionSyrtisMajor},
shows that orthopyroxene, clinopyroxene, olivine, goethite, clays
and maghemite are detected, in agreement with previous detection (\citep{Mustard_OlivinePyroxen_science_2005,Combe_MELSUM_PSS2008,Ehlmann_in-siturecordof_GRL2012}).
The fit of the spectra with highest mixing coefficient are available
in supplementary material. The estimated noise standard deviation
is $3.3\times10^{-4}$ from dark current but $5.0\times10^{-3}$ using
the MNF shift difference from ENVI software. The lack of fit estimated
by the RMS error is $2.1\times10^{-3}$ , in agreement with the previous
two extreme values. Using the fast GPU implementation, the computation
time is only 4 minutes for the complete OMEGA image of $M=$128x366
pixels, $N_{\lambda}=110$ bands. The computation time can can be
reduced to 0.3 min without any significant change in case of a recalibration
with $C\times100$.

\subsection{CRISM}

We propose to use the Compact Reconnaissance Imaging Spectrometer
for Mars (CRISM) onboard Mars Reconnaissance Orbiter (MRO) image frt0000A09C
(\citep{Murchie_CompactReconnaissanceImaging_JGR2007}) of Nili
Fossae where carbonate has been detected using manual band ratio's
technique on stacked denoise spectra, divided by a spectrally flat
component (\citep{Ehlmann_NiliFossae_JGR2009}). The data cube
has been radiometrically calibrated and the atmospheric gas transmission
has been empirically corrected using the volcano scan method (\citep{McGuire_CRISMatmocorrection_TGRS2008}).
We estimate the noise covariance matrix with the SNR=400 from the
calibration at the ground (\citep{Murchie_CompactReconnaissanceImaging_JGR2007})
in the diagonal elements. We also add a component in the non-diagonal
terms of the covariance matrix in order to take into account the correlation
of the bands due to atmospheric gas residuals. We corrected for the
striped noise by removing the small scale residual in each column
(\citep{2008LPI....39.2528P}). Nevertheless, this method is
not able to fully remove the non-gaussian noise and other more sophisticated
methods could be applied (\citep{2008LPI....39.2528P,Carter_Automatedprocessingplanetary_PaSS2013}).
We did not correct for spectral smile (\citep{Murchie_CompactReconnaissanceImaging_JGR2007})
unless this artifact can be corrected (\citep{Ceamanos_SpectralSmileCRISM_TGRS2010}).
Since the algorithm is comparing observed spectra to a single reference
spectra that cannot be shifted in wavelength, results may be affected
by the smile artifact. Nevertheless, the absorption bands of minerals
are often large and very insensitive to spectral smile, which is not
the case for gas and ices. Band ratios techniques may be less sensitive
to spectral smile for detection, but the value of the absorption depth
is also affected by the spectral smile.

\begin{figure}
\includegraphics[clip,width=1\columnwidth]{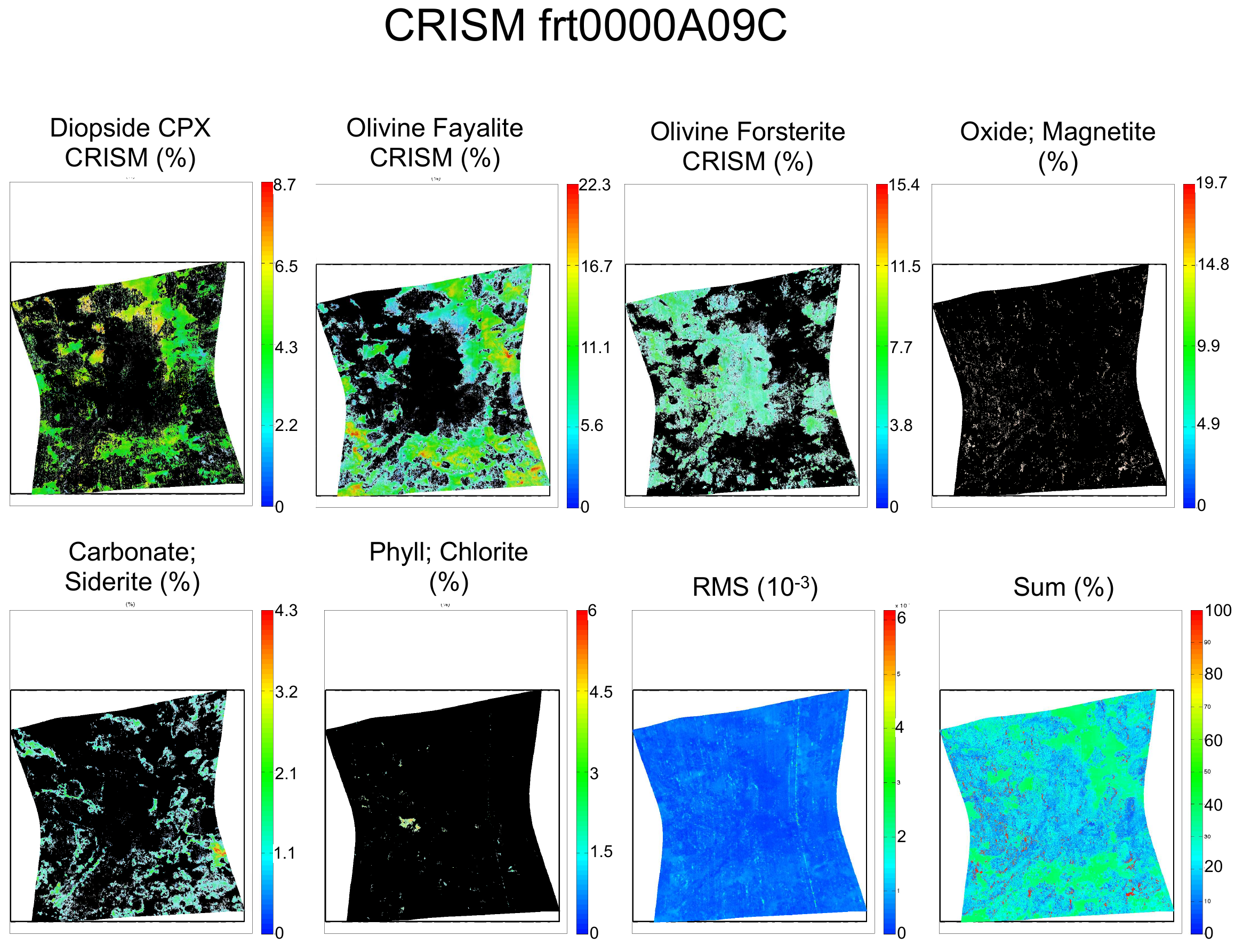}\centering\caption{Detection of 6 minerals over 36 spectra on CRISM image frt0000A09C
of Nili Fossae (21.3$^\circ$N, 78.5$^\circ$E) using IPLS
in the hue-saturation-value color system. The hue (color) represents
the mixing coefficient. The saturation (color or b/w) represents the
error. The value (intensity of color or b/w) represents the rms. Spectral
mixing coefficient map are shown with following conditions : (i) maximum
mixing coefficient $>$ 2\% , (ii) error on mixing coefficient $<$ mixing
coefficient, and(iii) RMS $<$ 10x the dark current noise (see text).
Pyroxenes, olivines, phyllosilicates and carbonate are detected
and the corresponding ``mixing coefficient" are mapped (color refer
to the online version of the article ).}
\label{fig:CRISMDetection} 
\end{figure}

\begin{figure}
\includegraphics[clip,width=1\columnwidth]{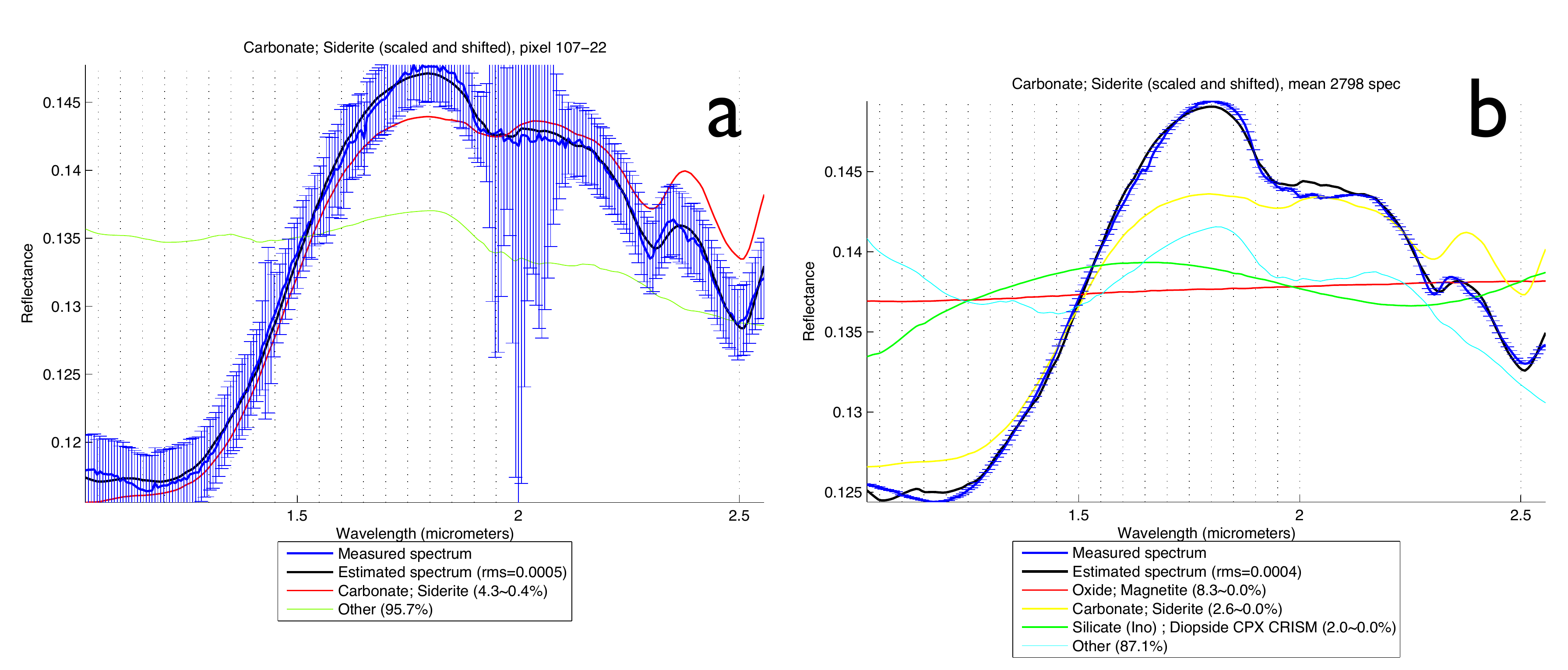}\centering

\caption{Example of spectra fit for the carbonate component (siderite in our
spectral database). All spectral endmember with significant mixing
coefficients ($>$2\%) are plotted but the other component are retrieved
by the algorithm (spectra called ``Other''), mainly due to atmospheric
transmission and water ice most probably in form of clouds. a) maximum
mixing coefficients spectra (pixel coordinates is line No 108, row
No 23), b) average value of 2896 spectra of the CRISM observation
frt0000A09C. \label{fig:Example-of-spectra-carbonate-CRISM}}
\end{figure}

The results in fig. \ref{fig:CRISMDetection} and \ref{fig:Example-of-spectra-carbonate-CRISM}
confirm the detection of carbonate of \citep{Ehlmann_NiliFossae_JGR2009}.
We stress the fact that our approach has been applied on raw image,
without stacking, neither dividing by a spectrally flat component,
in contrary to published detection.

The estimated noise standard deviation is $3.5\times10^{-4}$ from
SNR but $3.4\times10^{-3}$ using the MNF shift difference from ENVI
software. The lack of fit estimated by the RMS error is $1.0\times10^{-3}$
, in agreement with the previous two extreme values. Using the fast
GPU implementation, the computation time is 20 minutes for $M=$600x478
pixels, $N_{\lambda}=56$ bands. The computation time can can be reduced
to 2.7 min without any significant change in case of a recalibration
with $C\times100$.

The case of CRISM observation at a low SNR (frt0000a09) is exposed in 
supplementary material. No formal detection can be done because the structured 
noise level is higher than the signal level, with non-gaussian properties.

\subsection{M$^{3}$ }

We propose to use the Moon Mineralogy Mapper (M$^{3}$) onboard Chandrayaan-1
image G20090209T054031 (\citep{Pieters_M3_CurrentScience2009})
of Aristarchus, where peculiar olivines has been detected (\citep{LeMouelic_distributionolivinein_GRL1999,Chevrel_AristarchusPlateauMoon_I2009,Mustard_Aristarchus_JGR2011}).

\begin{figure}
\includegraphics[width=1\columnwidth]{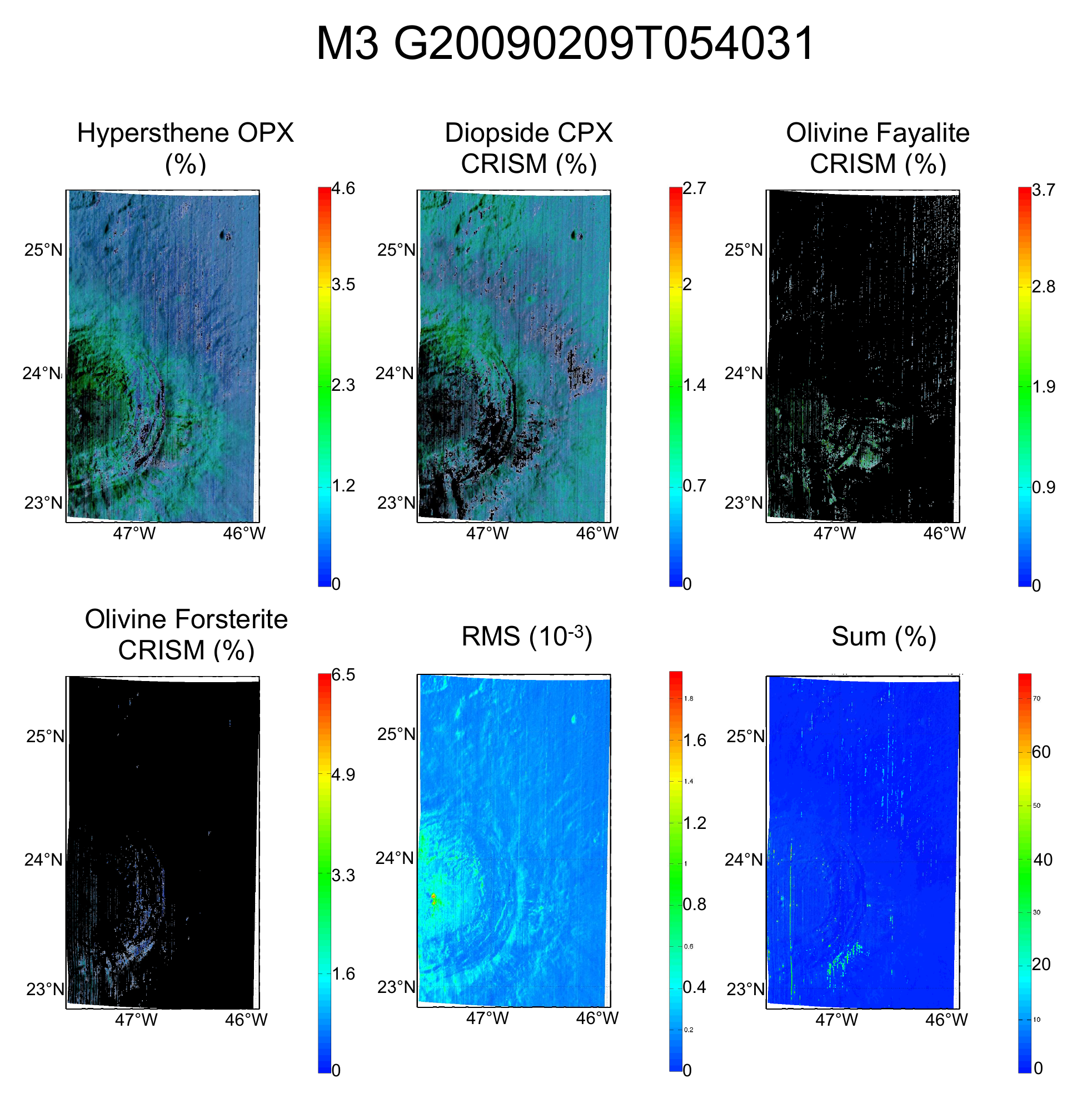}\centering\caption{Detection of 4 minerals over 36 spectra on M$^{3}$ image G20090209T054031
of Aristarchus using IPLS in the hue-saturation-value color system.
The hue (color) represents the mixing coefficient. The saturation
(color or b/w) represents the error. The value (intensity of color
or b/w) represents the rms. Spectral mixing coefficient map are shown
with following conditions : (i) maximum mixing coefficient $>$ 2\% ,
(ii) error on mixing coefficient $< $mixing coefficient, and (iii) RMS
$<$ 10x the dark current noise (see text). Pyroxenes and olivines are
detected and the corresponding ``mixing coefficient`` are mapped
(color refer to the online version of the article ).}
\label{fig:M3} 
\end{figure}

We estimate the SNR=400 from the calibration at the ground (\citep{Green_MoonMineralogyMapper_JGR2011}), 

The results in fig. \ref{fig:M3} also confirms that the South East
of Aristarchus is olivine rich but without any spectral signature
of pyroxene (\citep{LeMouelic_distributionolivinein_GRL1999,Chevrel_AristarchusPlateauMoon_I2009,Mustard_Aristarchus_JGR2011}).

The estimated noise standard deviation is $5.4\times10^{-5}$ from
SNR but $1.6\times10^{-3}$ using the MNF shift difference from ENVI
software. The lack of fit estimated by the RMS error is $2.1\times10^{-4}$,
in agreement with the previous two extreme values. Using the fast
GPU implementation, the computation time is 13 minutes for $M=$304x500
pixels, $N_{\lambda}=56$ bands due to the very small noise level
expected from the SNR ($5.4\times10^{-5}$ ). The computation time
can can be reduced to 1.6 min without any significant change in case
of a recalibration with $C\times100$.

\section{Conclusion}

We propose to incorporate additional spectra (constant, slope, and
cosine functions) to build a new supervised algorithm, called LinMin,
based on linear unmixing under positivity and sum-to-one constraints
in the goal to detect minerals at the planetary surfaces. The main
novelty of this approach is to treat ``linearly dependent
spectra" (such constant, slope and cosine functions)
that create degeneracies in usual unconstrained unmixing algorithms,
because under our constraints, they are not linearly dependent anymore.
Usually, the reference spectra database contains linearly independent
spectra (simply because they are different) so the degeneracies are
very small. Nevertheless, if the reference spectra database contains
significantly linearly dependent spectra, some degeneracies may occur.
Adding positivity constraint significantly reduces the degeneracies.
In the special case of linear subpixel mixing, the user may be interested
by the most precise aerial surface proportion. In this case, once
the detection step has been done, a second pass of the LinMin algorithm
with the detected spectrum only can enhance the surface proportion
estimation. 

Some hyperspectral images may be subject to ``spike'' noise, with
non-gaussian statistics. As stated in eq. \ref{eq:LinearMixing},
the linear unmixing methods usually assume gaussian noise, so the
results may be corrupted in the worst cases. One solution may be to
despike the data using average techniques (\citep{2008LPI....39.2528P,Carter_Automatedprocessingplanetary_PaSS2013})
before the treatment by LinMin. 

We validated the usefulness of LinMin to estimate mixing coefficients
in the case of linear mixture on synthetic examples. We also tested
numerically the most important non-linear effects on the detection
limits : aerosols content and grain size change. Both cases are well
treated with our modeling. Some minerals with large scale feature,
such olivines and pyroxenes, cannot be detected in the case of additional
cosine function because their spectral signature can be also fitted
with a mixture of cosine. Nevertheless, the use of additional spectra
made of constant and slope components improve the detection of all
minerals, including olivines and pyroxenes in comparison with no additional
spectra Our approach permits to save the continuum fitting step since
it is incorporated in the linear unmixing. We also showed that the
knowledge of the noise covariance matrix, that can be estimated from
dark current or using other techniques is important to assess the
detection limits, and in particular the error on mixing coefficients. 

We also tested LinMin on three real cases of hyperspectral images
from OMEGA, CRISM and M$^{3}$ instruments. All three cases show detections
in agreement with previous analysis, validating the LinMin with approach.
The main difficulty of this approach, that is also present in band
ratio and other detection method, is to optimize the threshold for
detection. Nevertheless, this problem is partly tackled by the estimation
of the error on the mixing coefficients. A combination of threshold
on RMS residues, error on mixing coefficients, and maximum mixing
coefficients seems to be the best compromise to ensure automatic detection.
This strategy has to be fully validated on large dataset.

IPLS is shown to be the best numerical algorithm to solve the LinMin
problem. Its fast GPU implementation is particularly relevant for
the treatment of large dataset of hyperspectral images. In the future,
this methodology should be applied in various planetary cases in order
to study the surface geology, especially in more challenging detection
situation such complex mafics and anorthosites assemblage on the Moon
(\citep{Ohtake_globaldistributionpure_N2009}), or mixture of
hydrous sulfates, hydrated acid and water ice on Europa (\citep{McCord_HydratedmineralsEuropa_I2010,Dalton_Exogeniccontrolssulfuric_PaSS2013}).
Also a significant improvement of the mineral detection may be addressed
by using spectral database adapted to the context.

\subsubsection*{Acknowledgement}

We acknowledge support from the ``Institut National des Sciences
de l'Univers'' (INSU), the ``Centre National de la Recherche
Scientifique" (CNRS) and ``Centre National
d'Etude Spatiale" (CNES) and through the "Programme
National de Plan{\'e}tologie".

\section*{Appendix}

Name of the 32 spectra: 

\begin{tabular}{|c|c|c|}
\hline 
{\scriptsize 1 Inosilicate (Hypersthene OPX PYX02.h >250u)} & {\scriptsize 12 Sulfate; Gypsum} & {\scriptsize 23 Carbonate; Siderite}\tabularnewline
\hline 
{\scriptsize 2 Inosilicate (Diopside CPX CRISM)} & {\scriptsize 13 Sulfate; Jarosite} & {\scriptsize 24 Phyllosilicate (Chlorite)}\tabularnewline
\hline 
{\scriptsize 3 Olivine Fayalite CRISM} & {\scriptsize 14 Sulfate; Kieserite} & {\scriptsize 25 Muscovite GDS116 Tanzania}\tabularnewline
\hline 
{\scriptsize 4 Olivine Forsterite CRISM} & {\scriptsize 15 Epsomite USGS GDS149} & {\scriptsize 26 Alunite GDS83 Na63}\tabularnewline
\hline 
{\scriptsize 5 Phyllosilicate (Clay Montmorillonite Bentonite)} & {\scriptsize 16 Oxide; Goethite} & {\scriptsize 27 Atmospheric Transmission}\tabularnewline
\hline 
{\scriptsize 6 Phyllosilicate (Clay Illite Smectite)} & {\scriptsize 17 Oxide; Hematite} & {\scriptsize 28 H2O grain 1}\tabularnewline
\hline 
{\scriptsize 7 Phyllosilicate (Serpentine Chrysotile Clinochry.)} & {\scriptsize 18 Oxide; Magnetite} & {\scriptsize 29 H2O grain 100}\tabularnewline
\hline 
{\scriptsize 8 Phyllosilicate (Serpentine Lizardite)} & {\scriptsize 19 Ferrihydrite USGS GDS75 Sy F6} & {\scriptsize 30 H2O grain 1000}\tabularnewline
\hline 
{\scriptsize 9 Phyllosilicate (Clay Illite)} & {\scriptsize 20 Maghemite USGS GDS81 Sy (M-3)} & {\scriptsize 31 CO2 grain 100}\tabularnewline
\hline 
{\scriptsize 10 Phyllosilicate (Clay Kaolinite)} & {\scriptsize 21 Carbonate; Calcite} & {\scriptsize 32 CO2 grain 10 000}\tabularnewline
\hline 
{\scriptsize 11 Phyllosilicate (Nontronite)} & {\scriptsize 22 Carbonate; Dolomite} & \tabularnewline
\hline 
\end{tabular}

Name of the 12 additional spectra: 

\begin{tabular}{|c|c|c|}
\hline 
33 Flat 1 & 37 cos 1/4 & 41 cos 1/2\tabularnewline
\hline 
34 Flat 0.0001 & 38 sin 1/4 & 42 sin 1/2\tabularnewline
\hline 
35 Slope Increasing & 39 -cos 1/4 & 43 -cos 1/2\tabularnewline
\hline 
36 Slope Decreasing & 40 -sin 1/4 & 44 -sin 1/2\tabularnewline
\hline 
\end{tabular}

\bibliographystyle{elsarticle-harv}

\end{document}